\documentclass[a4paper,12pt]{article}

\usepackage{amsmath,amssymb,booktabs}
\usepackage{epsf,epsfig}
\usepackage{psfrag}
\usepackage{cite}
\usepackage{graphics,subfigure}
\usepackage{mathrsfs}
\usepackage{bbm}

\usepackage{dcolumn}

\allowdisplaybreaks

\def\nnb{\nonumber}

\def\calH{\mathcal{H}}
\def\ub{\bar{u}}
\def\asfourpi{\frac{\alpha_s}{4\pi}}

\def\be{\begin{equation}}
\def\ee{\end{equation}}
\def\beq{\begin{eqnarray}}
\def\eeq{\end{eqnarray}}

\def\eps{\epsilon}

\def\slash#1{#1 \hskip-0.45em /}

\def\DB0{\partial B_0}

\def\Cl2{\mbox{Cl}_2}

\def\slash#1{#1 \hskip-0.45em /}

\usepackage{color}

\definecolor{Brown}{rgb}{0.5,0.25,0}

\begin{document}


\begin{titlepage}

\begin{flushright}
OUTP-15-15P\\
TUM-HEP-1005/15\\
SI-HEP-2015-17\\
QFET-2015-23\\[0.1cm]
July 13, 2015
\end{flushright}
\vskip 1.0cm

\begin{center}
\Large{\bf\boldmath
Two-loop current-current operator contribution to the
non-leptonic QCD penguin amplitude
\unboldmath}

\normalsize
\vskip 1.5cm

{\sc G.~Bell}$^{a}$,
{\sc M.~Beneke}$^{b}$,
{\sc T.~Huber}$^{c}$
and
{\sc Xin-Qiang~Li}$^{d,e}$\\

\vskip .5cm

{\it $^a$ Rudolf Peierls Centre for Theoretical Physics,
University of Oxford, 1 Keble Road, Oxford OX1 3NP,
United Kingdom
}\\[0.2cm]
{\it $^b$  Physik Department T31, Technische Universit\"at
M\"unchen, James-Franck-Stra\ss e~1, D-85748 Garching, Germany}\\[0.2cm]
{\it $^c$Theoretische Physik 1, Naturwissenschaftlich-Technische Fakult\"at,
\\ Universit\"at Siegen, Walter-Flex-Strasse 3, D-57068 Siegen, Germany}\\[0.2cm]
{\it $^d$Institute of Particle Physics and Key Laboratory of Quark and Lepton Physics~(MOE),
\\ Central China Normal University, Wuhan, Hubei 430079, P.~R. China}\\[0.2cm]
{\it $^e$State Key Laboratory of Theoretical Physics, Institute of Theoretical Physics,
\\ Chinese Academy of Sciences, Beijing 100190, P.~R. China}

\vskip 1.8cm

\end{center}

\begin{abstract}
\noindent
The computation of direct CP asymmetries in charmless $B$ decays
at next-to-next-to-leading order (NNLO) in QCD is of interest to
ascertain the short-distance contribution. Here we compute the
two-loop penguin contractions of the current-current operators
$Q_{1,2}$ and provide a first estimate of NNLO CP asymmetries in penguin-dominated $b\to s$ transitions.
\end{abstract}

\vfill

\end{titlepage}


\section{Introduction}
\label{sec:intro}

Non-leptonic exclusive decays of $B$ mesons play a crucial role in
studying the CKM mechanism of quark flavour mixing and in quantifying the
phenomenon of CP violation. Direct CP violation is related to the rate
difference of $\bar B\to f$ decay and its CP-conjugate and
arises if the decay amplitude is composed of at least two partial amplitudes
with different re-scattering (``strong'') phases, which are multiplied by
different CKM matrix elements. Very often useful information on the CKM
parameters including the CP-violating phase can be obtained from combining
different decay modes, whose partial amplitudes are related by the
approximate flavour symmetries of the strong
interaction~\cite{Zeppenfeld:1980ex}, which are then determined from data.

The direct computation of the partial amplitudes is a complicated strong
interaction problem, which can, however, be addressed in the heavy-quark
limit. The QCD factorization
approach~\cite{Beneke:1999br,Beneke:2000ry,Beneke:2001ev} employs
soft-collinear factorization in this limit to express the hadronic matrix
elements in terms of form factors and convolutions of perturbative objects
(hard-scattering kernels) with non-perturbative light-cone distribution
amplitudes (LCDAs). At leading order in $\Lambda/m_b$,
\begin{eqnarray}
\label{factformula}
\langle M_1 M_2 | Q_i | \bar{B} \rangle & = &
i m_B^2 \,\bigg\{f_{+}^{BM_1}(0)
\int_0^1 \! du \; T_{i}^I(u) \, f_{M_2}\phi_{M_2}(u)
+ (M_1\leftrightarrow M_2) \nnb \\
&& \hspace*{-1.5cm}
+ \,\int_0^\infty \! d\omega \int _0^1 \! du dv \;
T_{i}^{II}(\omega,v,u) \, \hat f_B \phi_B(\omega)  \; f_{M_1}\phi_{M_1}(v) \;
f_{M_2} \phi_{M_2}(u) \bigg\},
\end{eqnarray}
where $Q_i$ is a generic operator from the effective weak
Hamiltonian. At this order the re-scattering phases are generated at the
scale $m_b$ only, and reside in the loop corrections to the
hard-scattering kernels. Beyond the leading order factorization does not
hold, and re-scattering occurs at all scales. The leading contributions
to the strong phases are therefore of order $\alpha_s(m_b)$ or/and
$\Lambda/m_b$. It is of paramount importance for the predictivity
of the approach for the direct CP asymmetries to know whether the
short-distance or long-distance contribution dominates in practice, since
apart from being parametrically small, both could be numerically of similar
size.

The short-distance contribution to the direct CP asymmetries is fully
known only to the first non-vanishing order (that is,
${\cal O}(\alpha_s)$) through the
one-loop computations of the vertex kernels $T^{I}_i$ performed
long ago \cite{Beneke:1999br,Beneke:2001ev,Beneke:2003zv}. A reliable
result presumably requires the next-to-next-to-leading order
${\cal O}(\alpha_s^2)$ hard-scattering kernels, at least their imaginary
parts. For the spectator-scattering kernels $T_i^{II}$ this task is already
completed, both for the tree~\cite{Beneke:2005vv,Kivel:2006xc,Pilipp:2007mg}
and penguin~\cite{Beneke:2006mk,Jain:2007dy} amplitudes, but for the
so-called form factor term(s) in the first line of (\ref{factformula})
an important piece is still missing, which is the focus of this Letter.

We recall that due to CKM unitarity, the amplitude for a $\bar B$ decay
governed by the $b\to D$ ($D=d,s$) transition can always be written
in the form
\begin{equation}
\label{generalamplitude}
A(\bar B\to f) = \lambda_u^{(D)} \,[T+\ldots] +
\lambda_c^{(D)} \,[P_c +\ldots],
\end{equation}
where $\lambda_p^{(D)} = V_{pD}^* V_{pb}$. It is generic that the
first CKM structure is dominated by the colour-allowed or colour-suppressed
topological tree amplitude, both denoted by $T$ here,
corresponding to the flavour quantum numbers of a $b\to u\bar u D$
transition, while the second is dominated by the topological QCD penguin
amplitude of the $b\to \sum_{q=u,d,s} q\bar q D$ transition. The first is
typically larger than the second for $D=d$ and vice-versa
for the $D=s$ case, which therefore refers to the penguin-dominated
decays such as $\bar B\to \pi K$ and related. In the notation
of \cite{Beneke:2003zv,Beneke:2006mk}, $P_c$ corresponds to the
quantity $\alpha_4^c(M_1 M_2)$.\footnote{$\alpha_4^u(M_1 M_2)$ refers to a
generically sub-leading penguin contribution to the term multiplied by the
CKM factor~$\lambda_u^{(D)}$. We also note that $\alpha_4^p(M_1 M_2)$
consists of a leading-power term $a^p_4$ and a power-suppressed
term $a^p_6$~\cite{Beneke:2003zv}. The calculation reported
here concerns the leading-power contribution $a^p_4$.}

The vertex kernels $T_i^{I}$ have been computed at the two-loop
${\cal O} (\alpha_s^2)$ order only for the topological tree
amplitudes $T$~\cite{Bell:2007tv,Bell:2009nk,Beneke:2009ek}. However,
direct CP asymmetries can only be non-zero due to the interference
of the two terms in (\ref{generalamplitude}), hence the penguin
amplitude $P_c$ is also needed. Only the one-loop ${\cal O} (\alpha_s^2)$
contribution from the chromomagnetic dipole operator $Q_{8g}$ to $P_c$ has
been considered in the past~\cite{Kim:2011jm}, while the dominant, genuine
two-loop contributions remain to be computed. This calculation is technically
very involved since it requires the computation of massive two-loop penguin
diagrams -- a genuine two-loop, two-scale problem. One step towards this goal
was recently achieved in~\cite{Bell:2014zya}, where analytic results of all
occurring master integrals were derived.

At this point it is important to note that the topological tree and penguin
amplitudes are not in one-to-one correspondence with the tree (or
current-current) operators $Q_{1,2}^p$ and QCD penguin operators $Q_{3-6}$ of
the weak effective Hamiltonian. By contracting the $p\bar p$ fields of the
operators $Q_{1,2}^p$ (see (\ref{eq:Q12p}) below), they contribute to the QCD
penguin amplitude starting from the one-loop order. Since these ``penguin
contractions'' of the current-current operators come with the large
short-distance coefficients $C_{1,2}$, we may argue that they constitute the
largest contribution to the penguin amplitude at any given loop
order.\footnote{Since the contribution from
$Q_{1,2}^p$ alone is not renormalization-group invariant, this statement
cannot be true in arbitrary schemes nor at arbitrary renormalization
scales. What we mean is that the statement holds in the conventional
$\overline{\rm MS}$ scheme and with a reasonable choice ${\cal O}(m_b)$ of
scale.} At {\em next-to-leading order} we find for the penguin contractions
(including the chromomagnetic dipole operator $Q_{8g}$)
\begin{eqnarray}
a_4^u(\pi \bar K)_{|_{\rm NLO}} &=& (-0.0087 - 0.0172 i)_{|_{Q_{1,2}}} +
(0.0042 + 0.0041 i)_{|_{Q_{3-6}}}  + 0.0083_{|_{Q_{8g}}},
\nonumber \\[0.1cm]
a_4^c(\pi \bar K)_{|_{\rm NLO}} &=& (-0.0131 - 0.0102 i)_{|_{Q_{1,2}}}
+ (0.0042 + 0.0041 i)_{|_{Q_{3-6}}}  + 0.0083_{|_{Q_{8g}}},
\label{eq:a4}
\end{eqnarray}
where we separated the contributions from the current-current and the other
operators. While there is a cancellation for the real part, the imaginary
part from $Q_{1,2}^p$ is clearly dominant. If we add the vertex contractions
at leading (LO) and next-to-leading order (NLO) and consider the entire form
factor contribution to $a_4^p(M_1 M_2)$ at NLO, the second term changes to
$(-0.0266 + 0.0032 i)_{|_{Q_{3-6}}}$ in both expressions, and the imaginary
part from $Q_{1,2}^p$ is still by far dominant. Thus, at NLO, the
short-distance direct CP asymmetries are mainly determined by the one-loop
penguin contractions of the current-current operators. It is reasonable to
assume that the insertion of $Q_{1,2}^{u,c}$ at two loops also captures the
bulk of the yet unknown NNLO form factor contribution $T_i^I$ to the penguin
amplitudes $a_4^{u,c}$. In this Letter we report the result of this
computation together with a few phenomenological implications. We shall
provide more technical details together with the remaining contributions
from the QCD penguin operators $Q_{3-6}$, which require additional work on
infrared subtractions not present for $Q_{1,2}^p$, in a future publication.


\section{Outline of the calculation}
\label{sec:calculation}

The effective weak Hamiltonian for $b\to D$ transitions ($D=d,s$) is given by
\begin{equation}
\calH_\text{eff} =
   \frac{4G_F}{\sqrt{2}} \sum_{p=u,c} V_{pD}^* V_{pb}
   \bigg( C_1 Q_1^p + C_2 Q_2^p + \ldots\bigg)
   + \text{h.c.}.
\end{equation}
Here and in the following we give explicitly only the definitions
pertinent to the current-current operators relevant to our calculation.
We adopt the Chetyrkin-Misiak-M\"unz (CMM) operator
basis~\cite{Chetyrkin:1997gb}, where the
current-current operators are defined as
\begin{equation} \label{eq:Q12p}
Q_1^p = (\bar p_L \gamma^\mu T^A b_L )\,(\bar D_L \gamma_\mu T^A p_L),
\qquad
Q_2^p = (\bar p_L \gamma^\mu b_L)\, (\bar D_L \gamma_\mu p_L),
\end{equation}
in terms of left-chiral quark fields $q_L = \frac12(1-\gamma_5)q$.
In dimensional regularization the operator basis has to be supplemented by
evanescent (vanishing in $D=4$ dimensions)
operators, for which we adopt the convention of~\cite{Gorbahn:2004my}.

\begin{figure}[t]
\centerline{\includegraphics[width=.9\textwidth]{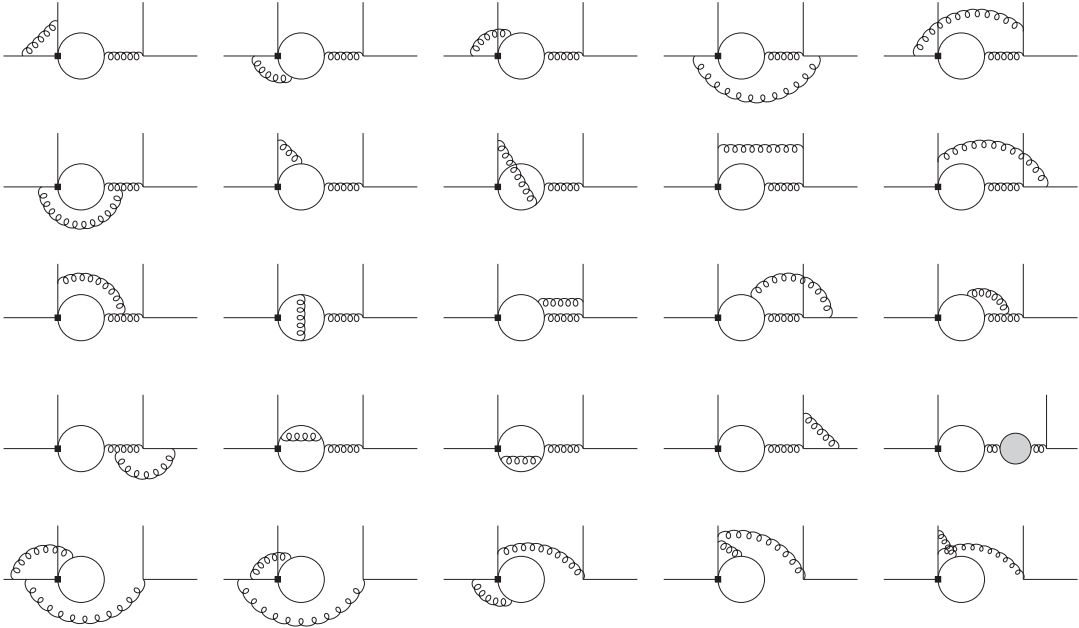}}
\caption{\label{fig:diagsQ12} Two-loop penguin diagrams that contribute
to the insertion of the operators $Q_{1,2}^{u,c}$ (black square). The grey
filled circle denotes the one-loop gluon self-energy.}
\end{figure}

At the two-loop level about $70$ diagrams contribute to the
QCD penguin amplitude, but only a subset of two dozens (shown in
Fig.~\ref{fig:diagsQ12}) are non-vanishing for the insertion of the
current-current operators $Q_{1,2}^p$. The quark in the fermion loop can
either be massless (for $p=u$) or
massive (for $p=c$). In the massless case the problem involves one
non-trivial scale, the momentum fraction $\bar u= 1-u$ of the anti-quark
in meson $M_2$, and the structure is similar to the NNLO calculation
of the tree amplitudes~\cite{Beneke:2009ek,Bell:2007tv,Bell:2009nk}. In the
massive case, however, we are dealing with a genuine two-loop, two-scale
problem since the hard-scattering kernels depend in addition on the mass
ratio $s_c=m_c^2/m_b^2$. As we have already elaborated extensively on the
kinematics in~\cite{Bell:2014zya}, we shall not repeat those formulae here.

The calculation is performed in dimensional regularization
with $D=4-2\eps$, where ultraviolet (UV) and infrared (soft and collinear)
divergences manifest themselves as poles in $\eps$. The CMM basis ensures
that the NDR scheme with a fully anti-commuting $\gamma_5$ can be adopted.
The amplitude of the diagrams is then computed using standard multi-loop
techniques. After a Passarino-Veltman~\cite{Passarino:1978jh} decomposition
of the tensor integrals, the scalar integrals are reduced to a
small set of master integrals by means of integration-by-parts
techniques~\cite{Tkachov:1981wb,Chetyrkin:1981qh} and the Laporta
algorithm~\cite{Laporta:1996mq,Laporta:2001dd}.
To this end, we use the program AIR~\cite{Anastasiou:2004vj} as well as
an in-house routine.

For the massless up-type operator insertions, the diagrams can be expressed
in terms of the master integrals that appeared in our former
calculations~\cite{Bell:2007tv,Bell:2009nk,Beneke:2009ek}. For the massive
charm-type insertions, on the other hand, we find 29 new master integrals.
The computation of the master integrals constitutes the main technical
challenge of the calculation. Analytic results for all master integrals have
recently been derived in~\cite{Bell:2014zya}, based on a differential
equation approach in a canonical basis~\cite{Henn:2013pwa}. The canonical
basis, together with suitably chosen kinematic variables, also catalyses the
convolution of the hard-scattering kernels with the LCDA.

After the computation of the bare QCD two-loop amplitude, the hard-scattering
kernels are extracted from a matching calculation onto
soft-collinear effective theory (SCET). The main conceptual challenge in this
context is the consistent treatment of evanescent and
Fierz-equivalent operators in SCET, for which we follow the method
employed in~\cite{Beneke:2009ek}. The SCET operators have the flavour
structure $\sum_q (\bar \chi_D \chi_q)(\bar\xi_q h_v)$ where $\chi$ and
$\xi$ denote collinear light-quark fields moving in opposite
directions. The two-loop diagrams relevant to the penguin amplitude $a_4^p$
are all of the ``wrong-insertion'' type (see~\cite{Beneke:2009ek}) and hence
lead to operators where the fermion indices are contracted in the form
$\sum_q (\bar \xi_q \chi_q)(\bar\chi_D h_v)$.
In Fig.~\ref{fig:diagsQ12} the $(\bar\xi_q \chi_q)$ fermion lines correspond
to the solid line on the right side of the diagram. In the following we
omit the sum over $q$ and the flavour labels on the fields.
In the CMM basis the fermion line that corresponds to
$(\bar \xi \chi)$ carries no $\gamma_5$ matrix. This
suggests that we use the following basis for the SCET operators:
\begin{align}
O_1 &= \bar \chi \, \frac{\slash{n}_{-}}{2} (1-\gamma_5) \chi \;
\bar \xi \, \slash{n}_{_+} (1-\gamma_5) h_v \, ,
\nonumber\\
\tilde O_n &=
\bar \xi \,
\gamma_{\perp}^{\alpha}\gamma_{\perp}^{\mu_1}\gamma_{\perp}^{\mu_2}
\ldots \gamma_{\perp}^{\mu_{2 n-2}} \chi \;
\bar \chi (1+\gamma_5)
\gamma_{\perp\alpha}\gamma_{\perp\mu_{2 n-2}}\gamma_{\perp\mu_{2n-3}}
\ldots\gamma_{\perp\mu_1} h_v,
\label{eq:scetbasis2}
\end{align}
where we need $n$ up to 2 (strings with three $\gamma$ matrices in
each bilinear). The operator $O_1$ is the only physical SCET operator.
It is the same as in \cite{Beneke:2009ek},
whereas the $\tilde O_n$ differ by the absence of the $1-\gamma_5$
factor to the left of $\chi$. The operators $\tilde O_n$ are evanescent
for $n>1$. $\tilde O_1$ is Fierz-equivalent to $O_1/2$ in four
dimensions, so we add $\tilde O_1-O_1/2$ as another evanescent
operator. We also recall that the SCET operators are
non-local on the light-cone \cite{Beneke:2009ek}.

After operator matching the hard-scattering kernels follow from the
bare QCD amplitudes plus subtraction terms from UV counterterms
of the operators $Q_i$ and the SCET operators. The master formulae
at LO, NLO, and NNLO read, respectively,
\begin{align}
\widetilde T_i^{(0)} &= \widetilde A^{(0)}_{i1} \, , \\
\widetilde T_i^{(1)} &= \widetilde A^{(1){\rm nf}}_{i1}+ Z_{ij}^{(1)} \,
\widetilde A^{(0)}_{j1}
 +\ldots\, , \label{eq:master1loopWI} \\
\widetilde T_i^{(2)} &= \widetilde A^{(2){\rm nf}}_{i1} +
Z_{ij}^{(1)} \, \widetilde A^{(1)}_{j1} + Z_{ij}^{(2)} \,
\widetilde A^{(0)}_{j1}
+ Z_{\alpha}^{(1)} \, \widetilde A^{(1){\rm nf}}_{i1}
+ \, (-i) \, \delta m^{(1)} \, \widetilde A^{\prime (1){\rm nf}}_{i1}\nnb \\
& + Z_{ext}^{(1)} \, \big[\widetilde A^{(1){\rm nf}}_{i1}+ Z_{ij}^{(1)} \,
\widetilde A^{(0)}_{j1}\big] - \,\widetilde T_i^{(1)} \big[  C_{FF}^{(1)}
+ \widetilde Y_{11}^{(1)}\big] +\ldots\, .
\label{eq:master2loopWI}
\end{align}
The symbols have the same meaning as in Eq.~(29) of~\cite{Beneke:2009ek}.
The ellipses denote further terms that do not contribute to the kernels
$i=1,2$ of the current-current operators. The matrices $Z_{ij}^{(1)}$ and
$Z_{ij}^{(2)}$ contain the UV counterterms from operator mixing. Compared
to the calculation of the tree amplitudes, they have to be extended by the
mixing with the penguin operators including the correspondent evanescent
operators~\cite{Gorbahn:2004my}. This implies, in particular, that the
one-loop amplitudes $\widetilde A^{(1)}_{j1}$ must be computed including
the $\cal{O}(\epsilon)$ terms for all operators $Q_j$, which mix with
the current-current operators. Finally, one has to convolute the
hard-scattering kernels with the LCDA, for which we adopt the conventional
Gegenbauer expansion.


\section{The topological QCD penguin amplitude}
\label{sec:amplitudes}

In this section we give the numerical results of the penguin
amplitudes $a_4^u$ and $a_4^c$ and discuss the size and scale dependence
of the new contribution. At LO, the penguin amplitude coefficients
are given in the CMM basis by ($N_c=3$, $C_F=4/3$)
\begin{equation}
a_{4,\rm LO}^p = \frac{1}{N_c}\left[C_3+C_F C_4+16 (C_5+C_F C_6)\right].
\end{equation}
They are identical for $p=u,c$ and independent of the LCDA. At NLO we
have ($L=\ln \mu^2/m_b^2$,
$s_p=m_p^2/m_b^2$, $\bar u=1-u$)
\begin{equation}
a_{4, {\rm NLO}}^p{|_{\,C_{1,2}}} = \asfourpi \frac{C_F}{N_c}
\left( C_2 - \frac{C_1}{2N_c} \right)
\int_0^1 du \,\left[ -\frac23L + \frac23 - G(s_p-i\epsilon,\ub) \right]
\phi_{M_2}(u),
\end{equation}
where we show only the terms from the current-current operators to illustrate
the structure of the result. Here
\begin{align}
G(s,u) &=  \frac{2 (12 s + 5 u - 3 u \ln s)}{9u} -
 \frac{2 \xi (2 s + u)}{3u} \ln \frac{\xi + 1}{ \xi - 1}
\end{align}
is the one-loop penguin function with $\xi = \sqrt{ 1- 4s/u }$.
In practice, one then inserts the Gegenbauer expansion of $\phi_{M_2}(u)$
truncated at the second order to perform the integral. The result is
finally expressed in terms of Wilson coefficients, quark masses and
the Gegenbauer moments $a^{M_2}_{1,2}$.

\begin{figure}[t]
\centerline{
\includegraphics[width=8cm]{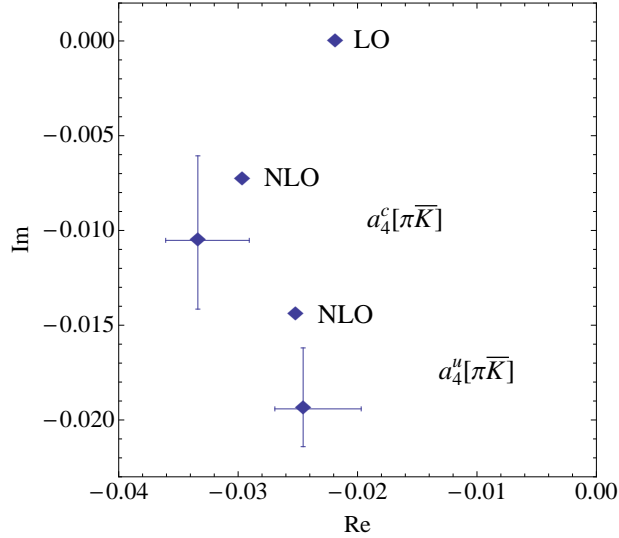}}
\caption{\label{a4plot} The LO, NLO and NNLO values of $a_4^{u}(\pi\bar K)$
and $a_4^{c}(\pi\bar K)$ in the complex plane. The NNLO point includes a
theoretical error estimate.}
\end{figure}

At NNLO the explicit expressions are involved, and we postpone
the details to a future publication.
Our final numerical predictions for the leading QCD penguin
amplitudes $a_4^{u,c}(\pi \bar K)$ are given as (for input parameters,
see section~\ref{sec:pheno}):
\begin{eqnarray}
a_4^u(\pi \bar{K})/10^{-2} &=& -2.87 -
[0.09 + 0.09i]_{\rm V_1} + [0.49-1.32i]_{\rm P_1} - [0.32+0.71i]_{\rm P_2}
\nonumber \\[0.2cm]
&& +\,\left[ \frac{r_{\rm sp}}{0.434} \right]
  \Big\{ [0.13]_{\rm LO} + [0.14 +0.12i]_{\rm HV} - [0.01-0.05i]_{\rm HP}
  + [0.07]_{\rm tw3} \Big\} \nonumber \\[0.2cm]
&=& (-2.46^{+0.49}_{-0.24}) + (- 1.94^{+0.32}_{-0.20})i\,,
\label{eq:a4unum} \\[1em]
a_4^c(\pi \bar{K})/10^{-2} &=& -2.87 -
[0.09 + 0.09i]_{\rm V_1} + [0.05-0.62i]_{\rm P_1} - [0.77+0.50i]_{\rm P_2}
\nonumber \\[0.2cm]
&& +\,\left[ \frac{r_{\rm sp}}{0.434} \right]
  \Big\{ [0.13]_{\rm LO} + [0.14 +0.12i]_{\rm HV} + [0.01+0.03i]_{\rm HP}
  + [0.07]_{\rm tw3} \Big\} \nonumber \\[0.2cm]
&=& (-3.34^{+0.43}_{-0.27}) + (-1.05^{+0.45}_{-0.36})i\,.
\label{eq:a4cnum}
\end{eqnarray}
In both equations the second line represents the spectator-scattering
term, which for $r_{\rm sp}=0.434$ makes only a small contribution to
$a_4^p$. In the respective first lines, the number without brackets is
the LO contribution, which has no imaginary part, the following two numbers
are the vertex and penguin NLO terms, and the new two-loop NNLO contribution
from the current-current operators $Q_{1,2}^p$ is the number labelled
$\rm P_2$. We observe that the new correction is rather large. It amounts
approximately to $40\%$~($15\%$) of the imaginary (real)
part of $a_4^u(\pi \bar K)$, and $50\%$~($25\%$) in the case of
$a_4^c(\pi \bar K)$. Graphical representations of $a_4^{p}(\pi \bar K)$
are shown in Fig.~\ref{a4plot} at LO, NLO and NNLO, where the NNLO point
includes the theoretical error estimate.\footnote{The LO and
NLO numbers here as in the subsequent figure are not the same
as (\ref{eq:a4unum}), (\ref{eq:a4cnum}) truncated to LO and NLO, because
they employ Wilson coefficients $C_i$
at LO and NLO, respectively. Moreover, consistent with previous LO and
NLO calculations, they are computed in the operator basis as defined
in \cite{BBL}. On the other hand, in (\ref{eq:a4unum}), (\ref{eq:a4cnum})
NNLO Wilson coefficients in the CMM basis are used throughout.}
The larger uncertainty of the
imaginary part of $a_4^c(\pi \bar K)$ is a consequence of the sensitivity
to the charm-quark (pole) mass, for which we adopt the conservative
range $m_c = 1.3\pm0.2\,$GeV.

\begin{figure}[t]
\centerline{\hskip-0.2cm
\includegraphics[width=15cm]{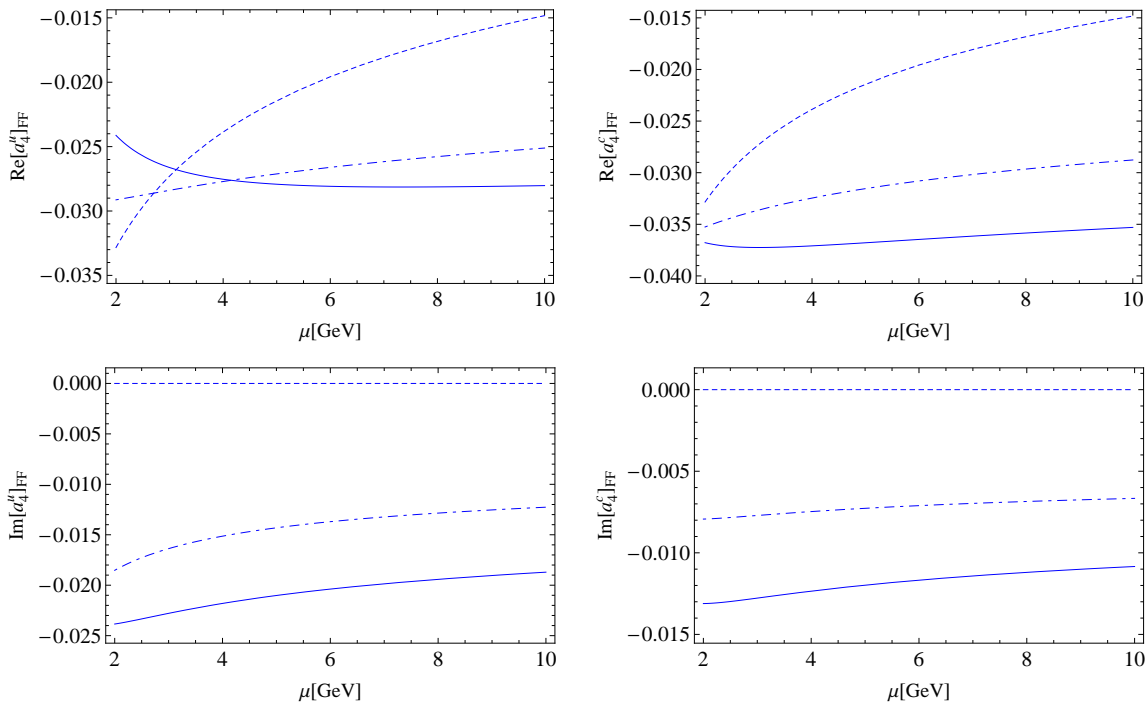}}
\caption{\label{fig:scale_dep_a4p} The dependence of the leading QCD
penguin amplitudes $a_4^{p}(\pi\bar K)$ on the hard renormalization scale
$\mu$~(form factor term only). Dashed, dashed-dotted and solid lines
represent LO, NLO, and NNLO, respectively.}
\end{figure}

The values (\ref{eq:a4unum}), (\ref{eq:a4cnum}) depend on the
renormalization scale due to the truncation of the perturbative expansion
and on hadronic parameters. The dependence on the renormalization scale $\mu$
may be considered as a measure of the accuracy of the approximation
at a given order in perturbation theory. This is shown in
Fig.~\ref{fig:scale_dep_a4p} for the form factor term contribution
to $a_4^p(\pi\bar K)$ up to NNLO. We observe a considerable stabilization
of the scale dependence for the real part, but less for the imaginary
part. This is explained by the fact that the imaginary part vanishes
at LO. Hence only the first correction is now available and is, moreover,
large.

\section{Phenomenology -- direct CP asymmetries}
\label{sec:pheno}

We now consider the new contribution to $a_4^p$ in the context of the
full QCD penguin amplitude and provide first results for some
direct CP asymmetries. We defer the discussion of branching fractions to
the more complete treatment including the two-loop matrix elements
of the penguin operators $Q_{3-6}$.

We recall that in the QCD factorization approach the full QCD penguin
amplitude consists of the parameters $a_4^p$, $a_6^p$, and
the penguin annihilation amplitude $\beta_3^p$ in the
combination~\cite{Beneke:2003zv}
\begin{equation}
\hat \alpha_4^p(M_1 M_2) = a_4^p(M_1 M_2) \pm r_\chi^{M_2} a_6^p(M_1 M_2)
+ \beta_3^p(M_1 M_2) ,
\label{a4hat}
\end{equation}
where the plus (minus) sign applies to the decays where $M_1$ is
a pseudoscalar (vector) meson. The first term, $a_4^p(M_1 M_2)$, is
the only leading-power contribution. Its real part is of order $-0.03$.
The annihilation term is $1/m_b$ suppressed and cannot be calculated
in the factorization framework. Estimates based on the model defined
in~\cite{Beneke:2001ev} suggest that its modulus is also of order
$0.03$. While the magnitude of these two terms is largely independent
of the spin of the final state mesons, the contribution from the
power-suppressed scalar penguin amplitude $r_\chi^{M_2} a_6^p(M_1 M_2)$
is very small when $M_2$ is a vector meson, but larger than the
leading-power amplitude for pseudoscalar $M_2$. It interferes constructively
for the $PP$ final state, and destructively for $VP$. It follows from
this brief discussion that the impact of a correction to $a_4^p$ is
always diluted in the full penguin amplitude. When $M_2=V$, the
computation of $a_4^p$ ascertains the short-distance contribution to the
amplitude, and hence the direct CP asymmetry, but there is an
uncertain annihilation contribution of similar size. When $M_2=P$, there
is another NNLO short-distance contribution from $a_6^p$, which is difficult
though not impossible to calculate, since it is power-suppressed.
These features will be clearly seen in the analysis below.

In the following we adopt the same values for the Standard
Model, meson and form factor parameters as in Table~1 of \cite{Beneke:2009ek}
with the exception of $|V_{ub}/V_{cb}| = 0.085 \pm 0.015$,
$\tau_{B_d}=1.52\,$ps, $m_s(2\,\mbox{GeV}) = (90\pm 10)\,$MeV, and
$f_{B_d} = (190\pm 10)\,$MeV. The decay constants, Gegenbauer moments
and form factors involving kaons coincide with~\cite{Beneke:2006mk},
those involving $K^\ast$ mesons
with~\cite{Beneke:2003zv}, except for $A_0^{BK^*}(0) = 0.39\pm 0.06$.
We note that the $B$-meson LCDA parameter $\lambda_B$ is not important
here, since the leading spectator-scattering contribution to the QCD
penguin amplitude is colour-suppressed.

\begin{figure}[p]
\centerline{
\includegraphics[width=15cm]{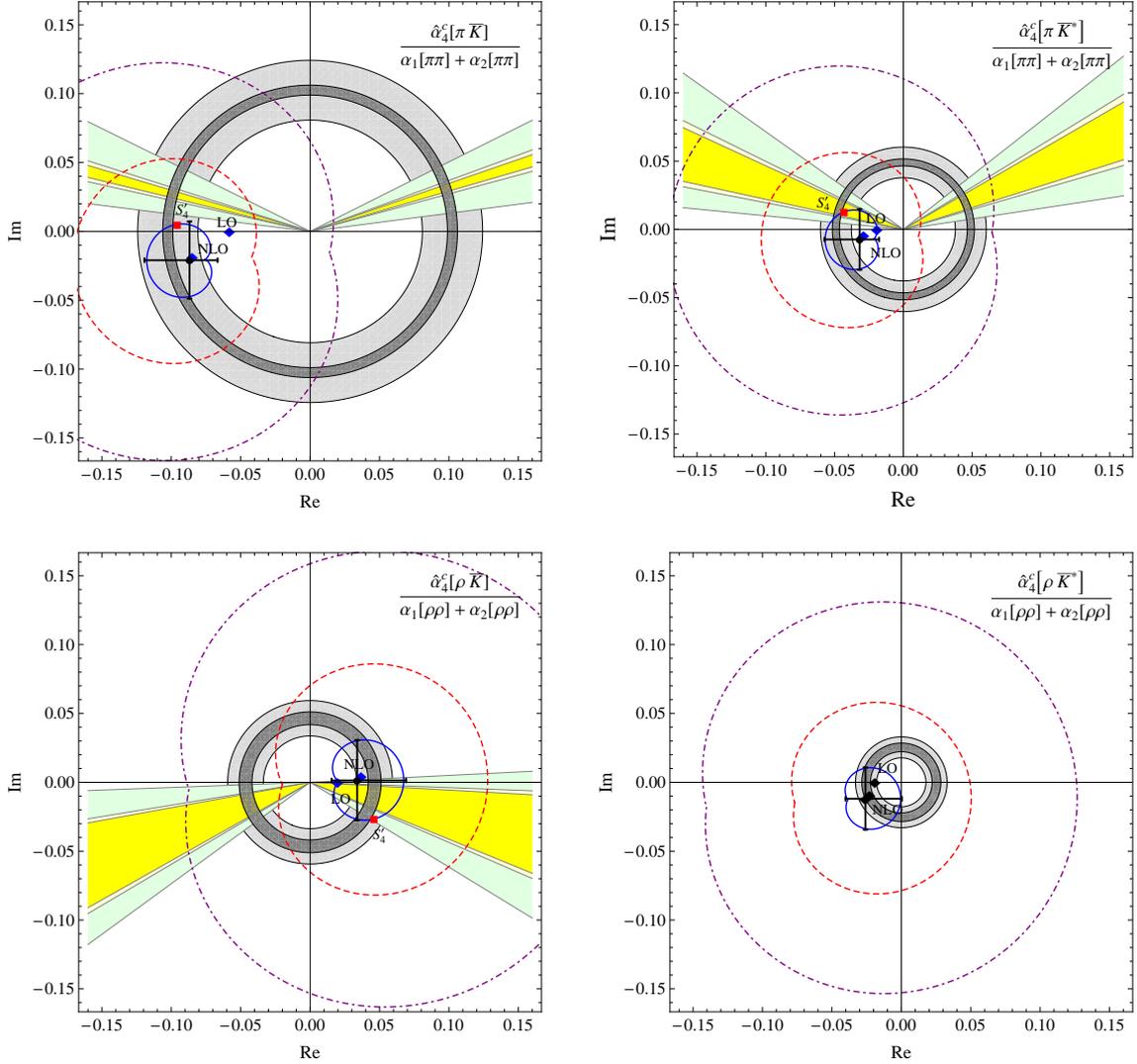}
}
\caption{\label{fig:penguinplot}
The QCD penguin amplitude $\hat \alpha_4^c(M_1 M_2)$ for the
$PP=\pi K$ final state and its $PV$, $VP$, and $VV$ relatives.
The $VV$ case refers to the longitudinal polarisation amplitude
only. Shown are the theoretical predictions for the ratios
$\hat \alpha_4^c(M_1 M_2)/(\alpha_1(\pi \pi) +
\alpha_2(\pi \pi))$ ($\rho\rho$ instead of $\pi\pi$ in the lower
row) and a comparison of extractions of the modulus (rings)
and phase (wedges) from data. Note there is no data for the
CP asymmetry in the rate of the longitudinally polarized
$\rho^+ K^{\ast -}$ final state.
See text for further explanations. }
\end{figure}
%

In Fig.~\ref{fig:penguinplot} we show the QCD penguin amplitude
$\hat \alpha_4^c(M_1 M_2)$ normalized to the sum of
colour-allowed and colour-suppressed tree amplitude $\alpha_1(\pi \pi) +
\alpha_2(\pi\pi)$,\footnote{For $M_1M_2=\pi \bar K,\, \pi \bar K^*$.
For $M_1M_2=\rho \bar K,\,\rho \bar K^*$ we use the $\rho\rho$ final state
instead. Also, for $\rho \bar K^*$ and $\rho\rho$,
only the longitudinal polarization amplitude
is considered in the following.} as was shown before
in~\cite{Beneke:2003zv,Beneke:2006mk}, but now includes the NNLO
computation for numerator and denominator. The NNLO result is represented
by the dark point with error bars and corresponds to setting
$\varrho_A=0$ in the annihilation model, which implies a small
value of $\beta_3^c$. The nearly circular contours
around this point show the variation of the theoretical prediction when
the phase of the annihilation model is varied from 0 to $2\pi$
for fixed $\varrho_A=1,2,3$ (inner to outer circles). The radius of
the circle for $\varrho_A=1$ leads to  the estimate
$|\beta_3^c|\approx 0.03$ mentioned above. The LO and NLO results
are marked by diamonds without error bars. Despite the
sizable NNLO correction to $a_4^c$ as shown in Fig.~\ref{a4plot},
the difference between NNLO and NLO is small. This is a consequence
of the ``dilution'' discussed above and a partial cancellation in the
ratio of amplitudes.

The theoretical prediction can be compared to data, since the
amplitude ratio can be related to CP-averaged decay rates $\Gamma$ and
direct CP asymmetries. We discuss this for the $PP$ case, from which
the others can be inferred by obvious replacements. First, to very
good approximation~\cite{Beneke:2003zv}
\begin{equation}
\label{a4piK}
  \left| \frac{\hat\alpha_4^c(\pi\bar K)}
              {\alpha_1(\pi\pi)+\alpha_2(\pi\pi)} \right|
  = \left| \frac{V_{ub}}{V_{cb}} \right| \frac{f_\pi}{f_K}
  \left[ \frac{\Gamma_{\pi^-\bar K^0}}
              {2\Gamma_{\pi^-\pi^0}} \right]^{1/2}\,,
\end{equation}
which determines the grey rings around the origin.
The darker rings are due to the experimental errors in
the branching fractions and the lighter ones include also the
uncertainty of $|V_{ub}/V_{cb}|$ (added linearly).
To obtain the wedges we define
$\psi$ to be the phase of the amplitude ratio shown in the figure, and
\begin{equation}
{\cal R} = \frac{\alpha_1(\pi\bar K)+\hat\alpha_4^u(\pi\bar K)}
{\alpha_1(\pi\pi)+\alpha_2(\pi\pi)}\,.
\end{equation}
We then find
\begin{eqnarray}
&& -\sin\psi + \frac{\mbox{Im}\,{\cal R}}{\mbox{Re}\,{\cal R}}
\,\cos\psi = \frac{1}{2\sin\gamma\,\mbox{Re}\,{\cal R}}\,
\left| \frac{V_{cs}}{V_{us}} \right| \frac{f_\pi}{f_K}
\, \frac{\Gamma_{\pi^+ K^-}}
{\sqrt{2 \Gamma_{\pi^-\pi^0}\Gamma_{\pi^-\bar K^0}}} \,A_{\rm CP}(\pi^+ K^-)
\,.
\label{wedge}
\end{eqnarray}
In previous discussions~\cite{Beneke:2003zv,Beneke:2006mk} the
experimental error on the observables on the right-hand side
and the error on $\gamma$ combined was large, so that it was
justified to assume that ${\cal R}$ is real and to neglect the
theoretical uncertainty on $\mbox{Re}\,{\cal R}$, which mainly stems
from the colour-suppressed tree amplitude $\alpha_2(\pi\pi)$.
This is no longer the case. The outer wedge now includes
the theoretical uncertainty on ${\cal R}$ and $\gamma$, which is
added linearly to the purely experimental uncertainties (inner wedge).
The middle wedge includes the uncertainty from $\gamma$ only.
Note that (\ref{wedge}) has two solutions as shown in the figure,
but the wedge that does not match the theoretical prediction is excluded
by $\Gamma_{\pi^+ K^-}/\Gamma_{\pi^- \bar K^0} <1$.

Since the NNLO correction to the amplitude ratio turned out to be small,
we can reaffirm the conclusions from \cite{Beneke:2006mk} in the light
of significantly improved data. The different magnitude of the $PP$
penguin amplitude vs. $PV$, $VP$ and $VV$ is clearly reflected in the
data as predicted. There is reasonable quantitative agreement as indicated
by the error bars and the small onion-shaped
regions corresponding to $\varrho_A=1$. An annihilation contribution
of $0.02$ to $0.03$ seems to be required, except for the longitudinal
$VV$ final states. The red square in the first three plots of
Fig.~\ref{fig:penguinplot} corresponds to the theoretical prediction
with $\varrho_A=1$ and the phase $\phi_A=-55^\circ\, (PP)$,
$\phi_A=-45^\circ\, (PV)$, $\phi_A=-50^\circ\, (VP)$
(see~\cite{Beneke:2001ev} for the definition of these quantities),
which is similar to
the favoured parameter set S4 of  \cite{Beneke:2003zv}. Only the
CP asymmetry of the $\pi K$ final state now appears to require
a value larger than $\varrho_A=1$ for a perfect fit.

\begin{table}[p]
\tabcolsep0.2cm
 \let\oldarraystretch=\arraystretch
 \renewcommand*{\arraystretch}{1.1}
\begin{center}
\begin{tabular}{lcccc}
\hline \hline
&&&&\\[-0.5cm]
$f$  & ${\rm NLO}$ & ${\rm NNLO}$ &  ${\rm NNLO}+{\rm LD}$ & Exp \\
\hline
&&&&\\[-0.3cm]
$\pi^-\bar{K}^0$
& $\phantom{-}0.71_{\,-0.14\,-0.19}^{\,+0.13\,+0.21}$
& $\phantom{-}0.77_{\,-0.15\,-0.22}^{\,+0.14\,+0.23}$
& $\phantom{-}0.10_{\,-0.02\,-0.27}^{\,+0.02\,+1.24}$
& $-1.7\pm1.6$ \\ \addlinespace
$\pi^0K^-$
& $\phantom{-}9.42_{\,-1.76\,-1.88}^{\,+1.77\,+1.87}$
& $10.18_{\,-1.90\,-2.62}^{\,+1.91\,+2.03}$
& $-1.17_{\,-0.22\,-\phantom{0}6.62}^{\,+0.22\,+20.00}$
& $\phantom{-}4.0\pm2.1$ \\ \addlinespace
$\pi^+K^-$
& $\phantom{-}7.25_{\,-1.36\,-2.58}^{\,+1.36\,+2.13}$
& $\phantom{-}8.08_{\,-1.51\,-2.65}^{\,+1.52\,+2.52}$
& $-3.23_{\,-0.61\,-\phantom{0}3.36}^{\,+0.61\,+19.17}$
& $-8.2\pm0.6$ \\ \addlinespace
$\pi^0\bar{K}^0$
& $-4.27_{\,-0.77\,-2.23}^{\,+0.83\,+1.48}$
& $-4.33_{\,-0.78\,-2.32}^{\,+0.84\,+3.29}$
& $-1.41_{\,-0.25\,-6.10}^{\,+0.27\,+5.54}$
& $\phantom{-1}1\pm10$ \\ \addlinespace
$\delta(\pi \bar{K})$
& $\phantom{-}2.17_{\,-0.40\,-0.74}^{\,+0.40\,+1.39}$
& $\phantom{-}2.10_{\,-0.39\,-2.86}^{\,+0.39\,+1.40}$
& $\phantom{-}2.07_{\,-0.39\,-4.55}^{\,+0.39\,+2.76}$
& $12.2\pm 2.2$ \\ \addlinespace
$\Delta(\pi \bar{K})$
& $-1.15_{\,-0.22\,-0.84}^{\,+0.21\,+0.55}$
& $-0.88_{\,-0.17\,-0.91}^{\,+0.16\,+1.31}$
& $-0.48_{\,-0.09\,-1.15}^{\,+0.09\,+1.09}$
& $-14\pm11$ \\ \addlinespace
\hline
&&&&\\[-0.3cm]
$\pi^-\bar{K}^{\ast0}$
& $\phantom{-1}1.36_{\,-0.26\,-0.47}^{\,+0.25\,+0.60}$
& $\phantom{-}1.49_{\,-0.29\,-0.56}^{\,+0.27\,+0.69}$
& $\phantom{-}0.27_{\,-0.05\,-0.67}^{\,+0.05\,+3.18}$
& $-3.8\pm4.2$ \\ \addlinespace
$\pi^0K^{\ast-}$
& $\phantom{1}13.85_{\,-2.70\,-5.86}^{\,+2.40\,+5.84}$
& $\phantom{-}18.16_{\,-3.52\,-10.57}^{\,+3.11\,+\phantom{0}7.79}$
& $-15.81_{\,-2.83\,-15.39}^{\,+3.01\,+69.35}$
& $\hspace*{0.1cm}-6\pm24$ \\ \addlinespace
$\pi^+K^{\ast-}$
& $\phantom{-}11.18_{\,-2.15\,-10.62}^{\,+2.00\,+\phantom{0}9.75}$
& $\phantom{-}19.70_{\,-3.80\,-11.42}^{\,+3.37\,+10.54}$
& $-23.07_{\,-4.05\,-20.64}^{\,+4.35\,+86.20}$
& $-23\pm6$ \\ \addlinespace
$\pi^0\bar{K}^{\ast0}$
& $-17.23_{\,-3.00\,-12.57}^{\,+3.33\,+\phantom{0}7.59}$
& $-15.11_{\,-2.65\,-10.64}^{\,+2.93\,+12.34}$
& $\phantom{-}2.16_{\,-0.42\,-36.80}^{\,+0.39\,+17.53}$
& $-15\pm13$ \\ \addlinespace
$\delta(\pi \bar{K}^{\ast})$
& $\phantom{-}2.68_{\,-0.67\,-4.30}^{\,+0.72\,+5.44}$
& $-1.54_{\,-0.58\,-9.19}^{\,+0.45\,+4.60}$
& $\phantom{-}7.26_{\,-1.34\,-20.65}^{\,+1.21\,+12.78}$
& $\phantom{-}17\pm 25$ \\ \addlinespace
$\Delta(\pi \bar{K}^{\ast})$
& $-7.18_{\,-1.28\,-5.35}^{\,+1.38\,+3.38}$
& $-3.45_{\,-0.59\,-4.95}^{\,+0.67\,+9.48}$
& $-1.02_{\,-0.18\,-7.86}^{\,+0.19\,+4.32}$
& $-5\pm45$ \\ \addlinespace
\hline
&&&&\\[-0.3cm]
$\rho^-\bar{K}^0$
& $\phantom{-}0.38_{\,-0.07\,-0.27}^{\,+0.07\,+0.16}$
& $\phantom{-}0.22_{\,-0.04\,-0.17}^{\,+0.04\,+0.19}$
& $\phantom{-}0.30_{\,-0.06\,-2.39}^{\,+0.06\,+2.28}$
& $-12\pm17$ \\ \addlinespace
$\rho^0K^-$
& $-19.31_{\,-3.61\,-\phantom{0}8.96}^{\,+3.42\,+13.95}$
& $-4.17_{\,-0.80\,-19.52}^{\,+0.75\,+19.26}$
& $\phantom{-}43.73_{\,-7.62\,-137.77}^{\,+7.07\,+\phantom{0}44.00}$
& $\phantom{-}37\pm11$ \\ \addlinespace
$\rho^+K^-$
& $-5.13_{\,-0.97\,-4.02}^{\,+0.95\,+6.38}$
& $\phantom{-}1.50_{\,-0.27\,-10.36}^{\,+0.29\,+\phantom{0}8.69}$
& $\phantom{-}25.93_{\,-4.90\,-75.63}^{\,+4.43\,+25.40}$
& $\phantom{-}20\pm11$ \\ \addlinespace
$\rho^0\bar{K}^0$
& $\phantom{-}8.63_{\,-1.65\,-1.69}^{\,+1.59\,+2.31}$
& $\phantom{-}8.99_{\,-1.71\,-7.44}^{\,+1.66\,+3.60}$
& $\phantom{1}-0.42_{\,-0.08\,-\phantom{0}8.78}^{\,+0.08\,+19.49}$
& $\phantom{-1}6\pm20$ \\ \addlinespace
$\delta(\rho \bar{K})$
& $-14.17_{\,-2.96\,-5.39}^{\,+2.80\,+7.98}$
& $-5.67_{\,-1.01\,\phantom{1}-9.79}^{\,+0.96\,+10.86}$
& $\phantom{-}17.80_{\,-3.01\,-62.44}^{\,+3.15\,+19.51}$
& $\phantom{-}17\pm 16$ \\ \addlinespace
$\Delta(\rho \bar{K})$
& $-8.75_{\,-1.66\,-6.48}^{\,+1.62\,+4.78}$
& $-10.84_{\,-2.09\,-\phantom{0}9.09}^{\,+1.98\,+11.67}$
& $\phantom{1}-2.43_{\,-0.42\,-19.43}^{\,+0.46\,+\phantom{0}4.60}$
& $-37\pm37$ \\ \addlinespace
\hline \hline
\end{tabular}
\end{center}
\caption{\label{tab:acp}
Direct CP asymmetries in percent for the $\pi K$, $\pi K^{\ast}$,
and $\rho K$ final states. The theoretical errors shown correspond
to the uncertainties due to the CKM and hadronic parameters,
respectively. The errors on the experimental values of $\delta$
and $\Delta$ are
computed from those of the individual observables appearing in
(\ref{eq:sumrule}) ignoring possible correlations.}
\end{table}

Moving to the observables themselves, we show in Table~\ref{tab:acp}
the theoretical predictions for direct CP asymmetries, defined as the
rate asymmetry between $\bar B$ and $B$ decays, together with
the world average of experimental data (last column), compiled from
HFAG~\cite{Amhis:2012bh}. We focus on the penguin-dominated $b\to s$
transitions of non-strange $B$ mesons to $\pi K$ final states and their
$PV$ and $VP$ relatives. We also show the CP asymmetry difference
\begin{equation}\delta(\pi \bar{K}) =
A_{\rm CP}(\pi^0 K^-) - A_{\rm CP}(\pi^+ K^-)
\end{equation}
and the
asymmetry ``sum rule''
\begin{equation}
\Delta(\pi \bar{K}) = A_{\rm CP}(\pi^+ K^-) +
\frac{\Gamma_{\pi^-\bar{K}^0}}{\Gamma_{\pi^+ K^-}} \,
A_{\rm CP}(\pi^- \bar{K}^0)
- \frac{2\Gamma_{\pi^0 K^-}}{\Gamma_{\pi^+ K^-}} \,
A_{\rm CP}(\pi^0 K^-)
- \frac{2 \Gamma_{\pi^0 \bar{K}^0}}{\Gamma_{\pi^+ K^-}} \,
A_{\rm CP}(\pi^0 \bar{K}^0)\,.
\label{eq:sumrule}
\end{equation}
The latter quantity is expected to be small \cite{Gronau:2005kz}, since
the leading CP-violating interference of QCD penguin and tree amplitudes
cancels out in the sum. In order to focus on the effect of the new NNLO
correction on the perturbatively calculable short-distance part of the
CP asymmetry, the columns labelled ``NLO'' and ``NNLO'' give the
respective results, when the long-distance, power-suppressed terms
are set to zero. This means that we set $\beta_3^p$ to zero, as well
as power-suppressed spectator-scattering terms. However, we keep the
short-distance dominated, but power-suppressed scalar penguin
contributions. The column labelled ``NNLO+LD'' adds the previously
neglected terms back. The main effect is from weak annihilation, for which
we adopt the S4-like scenario ($S_4^\prime$) marked by the red square in
Fig.~\ref{fig:penguinplot}.

Focusing first on the ``NLO'' and ``NNLO'' results, we note that
for the $PP$ final states the change is minor, since, as discussed above,
$a_4^c$ represents only part of the short-distance penguin amplitude.
The situation is different for the $\pi K^*$ final states where the $a_6^c$
contribution is small, and for
the $\rho K$ final states where due to the opposite
sign of $a_4^c$ and $a_6^c$ a cancellation occurs. In these cases, we
observe a large modification for the $\pi^0 K^{*-}$, $\pi^+ K^{*-}$
and the corresponding $\rho K$ final states, for which the CP asymmetry
arises predominantly from the imaginary part of $\hat\alpha_4^c/\alpha_1$.
These modifications are a reflection of the sizable corrections
seen in Fig.~\ref{a4plot}. The effect is much less pronounced in the
remaining modes, where the asymmetry is due to interference with
$\hat\alpha_4^u$ (in case of $\pi^- \bar{K}^{*0}$, $\rho^- \bar{K}^{0}$) or
$\alpha_2$ (in case of
$\pi^0 \bar{K}^{*0}$, $\rho^0 \bar{K}^{0}$), and the effect
of the NNLO correction cancels to a certain extent in the ratio of
interfering amplitudes. Despite these large modifications of some of
the $PV$ and $VP$ modes' asymmetries, the long-distance annihilation
contribution is always more important numerically, and usually required
to obtain a satisfactory description of the data. The modelling of
the long-distance contribution also determines the final theoretical
uncertainty, which can become very large. Given that the short-distance
contribution is now known to NNLO and given the large amount of
experimental data, it becomes imperative to better determine the
annihilation amplitude, presumably through fits to data.


\section{Conclusion}
\label{sec:conclusion}


The computation of direct CP asymmetries in charmless $B$ decays
at next-to-next-to-leading order in QCD has been a long-standing issue. The
long- and short-distance contributions can in principle be of the
same order and a NNLO calculation is required to ascertain the
perturbative part. In this paper we computed the two-loop contributions
of the current-current operators $Q_{1,2}^p$ to the QCD penguin amplitude,
which are expected to constitute the dominant contribution, at least
to the imaginary part, which is required for observing CP violation.
We find a sizable correction to the short-distance part of the
direct CP asymmetry, the effect of which is, however,
tempered by power-suppressed short- and long-distance terms. Our
preliminary conclusion is that the NNLO correction does not help
resolving the $\pi K$ CP asymmetry puzzle, nor does it render the
poorly known annihilation terms redundant. The final analysis should,
however,  include the penguin operator matrix elements, as well as the
one from the chromomagnetic operator considered in \cite{Kim:2011jm}.
The corresponding calculations are in progress.


\subsubsection*{Acknowledgements}
This work was supported in part by the NNSFC of China under
contract Nos.~11005032 and 11435003~(XL), by DFG Forschergruppe
FOR 1873 ``Quark Flavour Physics and Effective Field Theories''~(TH) and
by the DFG
Sonder\-for\-schungs\-bereich/Trans\-regio~9 ``Computergest\"utzte
Theoreti\-sche Teilchenphysik'' (MB). GB~gratefully acknowledges the
support of a University Research Fellowship by the Royal Society.
XL is also supported in part by the SRF for ROCS, SEM, by the Open
Project Program of SKLTP, ITP, CAS~(No.~Y4KF081CJ1), and by the
self-determined research funds of CCNU from the colleges' basic research
and operation of MOE~(CCNU15A02037). MB and XL acknowledge the
hospitality of the Munich Institute for Astro- and Particle Physics (MIAPP)
of the DFG cluster of excellence ``Origin and Structure of the Universe'',
where this work was finalized.



\end{document}